\documentclass[showpacs,preprintnumbers,amssymb]{revtex4}

\usepackage{graphicx,epsfig}
\usepackage{bm}

\newcommand{\be}{\begin{equation}}
\newcommand{\ee}{\end{equation}}
\newcommand{\bea}{\begin{eqnarray}}
\newcommand{\eea}{\end{eqnarray}}
\newcommand{\ba}{\begin{array}}
\newcommand{\ea}{\end{array}}
\newcommand{\beqn}{\begin{eqnarray*}}
\newcommand{\eeqn}{\end{eqnarray*}}

\newcommand{\f}[2]{\frac{#1}{#2}}

\def\nn{\nonumber}

\def\ii{{\rm i}}

\begin{document}

\nopagebreak

\title{Quasinormal modes of Kerr-Newman black holes: \\ coupling of
electromagnetic and gravitational perturbations} 

\author{Emanuele Berti}

\affiliation{\it McDonnell Center for the Space Sciences, 
Department of Physics, Washington University, 
St. Louis, Missouri 63130, USA}

\author{Kostas D. Kokkotas}

\affiliation{\it Department of Physics, Aristotle University of
Thessaloniki 54124, Greece}

\begin{abstract}
\noindent

We compute numerically the quasinormal modes of Kerr-Newman black
holes in the scalar case, for which the perturbation equations are
separable. Then we study different approximations to decouple
electromagnetic and gravitational perturbations of the Kerr-Newman
metric, computing the corresponding quasinormal modes. Our results
suggest that the Teukolsky-like equation derived by Dudley and Finley
gives a good approximation to the dynamics of a rotating charged black
hole for $Q\lesssim M/2$.  Though insufficient to deal with
Kerr-Newman based models of elementary particles, the Dudley-Finley
equation should be adequate for astrophysical applications.

\end{abstract}
\maketitle

\section{Introduction}
\label{sec:intro}

The electrovacuum black hole solutions of the Einstein-Maxwell system
can be uniquely described by the Kerr-Newman (KN) metric \cite{KN},
which is the most general among classical black hole solutions. The KN
metric is specified by three parameters: the black hole mass $M$, the
charge $Q$ and the angular momentum per unit mass $a\equiv J/M$. As
long as $M^2\geq Q^2+a^2$ the KN metric describes a black hole,
otherwise it has a naked ring-like singularity. As $Q\to 0$ the KN
metric reduces to the rotating Kerr metric, and as $a\to 0$ it reduces
to the Reissner-Nordstr\"om (RN) metric. Both limits have been studied
in great detail \cite{MTB,QNM}.

As early as 1968 Carter realized that the KN solution has a magnetic
dipole moment corresponding to the same g-factor $g=2$ as the electron
\cite{Carter}. This led to the suggestion (recently revisited in
\cite{Rosquist}: see discussion and references therein) that the KN
metric could provide a reasonably adequate model for the external
Einstein-Maxwell field of elementary particles \cite{KNmodels,pekeris}.

For astrophysical black holes $Q$ is likely to be negligible, electric
charge being shorted out by the surrounding plasma \cite{BZ}. If any
charge is present, the no-hair theorem guarantees that charged
rotating astrophysical black holes are described by the KN metric.
Punsly proposed a gamma-ray burst model based on bipolar outflow from
a fast-rotating black hole endowed with a small (by gravitational
standards) charge and surrounded by a magnetosphere \cite{P}. Van
Putten suggested that hypernovae or black hole-neutron star
coalescence may generate rapidly spinning black hole-torus systems,
and that the spin energy of the hole could power gamma-ray bursts
\cite{VP}. In his model the black hole-torus system has a
gravitationally weak magnetic field as a result of the remnant flux
from the progenitor star (a massive star in hypernovae or a neutron
star for a coalescence remnant), and the interplay between rotational
effects and magnetic fields may be relevant. Ruffini {\it et al.}
noticed that pair creation induced by the Heisenberg-Euler-Schwinger
vacuum polarization during collapse leading to formation of a charged
black hole could explain the energetics of gamma-ray bursts, and
proposed a detailed collapse model in a series of papers
\cite{ruffini}. More recently, Araya-Gochez suggested that
intermittent hyper-accretion produced by magnetorotational
instabilities in the accretion disc of a rapidly rotating, newly born
black hole may induce resonant excitation of the black hole's
quasinormal modes. His estimates suggest that a 15 $M_\odot$ black
hole spinning at $a\simeq 0.98 M$ and located at 27~Mpc could produce
gravitational waves detectable by LIGO II \cite{AG}.

Both in elementary particle models and in astrophysical applications
rotation plays a crucial role. In geometrical units, the typical
angular momentum of an electron $a\sim \hbar/2m_e=1.93\times
10^{-11}$~cm, the length associated to the electron charge
$Q=e=1.38\times 10^{-34}$~cm and the corresponding mass scale
$M=m_e=6.76\times 10^{-56}$~cm, so that $M\ll Q\ll a$ and the
gravitational field is spin-dominated. In fact, the rotation parameter
is so large that the KN metric can only model the Einstein-Maxwell
field of the electron outside some small radius $r_0$ surrounding the
ring-like naked singularity.
Astrophysical black holes are also expected to be formed in rapid
rotation: recent simulations suggest that supermassive stars would
form black holes with $a/M\sim 0.75$, and many observations are
consistent with black holes spinning close to the extremal limit
$a/M\sim 1$. Reference~\cite{shapiro} provides an updated discussion
of the observational evidence for black hole spins and of the related
uncertainties in present-day astrophysical models. Most black hole
models for gamma-ray bursts require rapid rotation \cite{VP,P}, and
the magnetorotational instability considered in \cite{AG} is most
efficient for large spin parameters, $a/M\gtrsim 0.9$.

The KN solution is the only asymptotically flat solution of the
Einstein-Maxwell system for which the geodesic and Klein-Gordon
equations can be solved by separation of variables \cite{DT}. The
Dirac equation in the KN metric is also known to be separable
\cite{page}. Scalar and Dirac perturbations of a KN black hole can
therefore be treated using the same general methods that apply to Kerr
black holes. In particular, it is straightforward to compute the
quasinormal modes (QNMs) of scalar perturbations of the KN black
hole. As far as we know, a complete analysis of the corresponding QNM
spectrum is still lacking. The first objective of this paper is to
fill this gap, presenting a complete continued-fraction calculation of
scalar QNMs of the KN metric for all values of $a$ and $Q$.

Studies of the interplay of electromagnetic (EM) and gravitational
fields in the KN metric are plagued by a major technical difficulty:
all attempts to decouple the EM and gravitational perturbations of the
KN spacetime to date have failed. Section 111 of \cite{MTB} gives an
introduction to this long-standing unsolved problem. Dudley and Finley
(\cite{DF}, henceforth DF; see also \cite{DFPRL}) make a remarkable
study of the separability of linear perturbations of the solutions of
the Einstein-Maxwell equations found by Pleba\'nski and Demia\'nski
\cite{PD}, which include all vacuum Type D solutions \cite{Weir}.  In
their work Dudley and Finley ``either keep the geometry fixed and
perturb the electric field or, of more interest, keep the electric
field fixed and perturb the geometry'' \cite{DF}. This approach should
be appropriate for values of the charge $Q$ at most as large as the
perturbations of the spacetime metric (in geometrized units). Dudley
and Finley show that a sufficient condition for decoupling is that the
spacetime be of Type D, and that the decoupled equations only separate
(in Pleba\'nski-Demia\'nski coordinates) for perturbing fields of spin
$s=0,~1/2,~1$ and $2$.

Mashhoon \cite{Mash} first presented arguments in favor of the
stability of the KN metric. Instead of explicitly computing QNMs, he
used an approximate argument (originally due to Goebel \cite{goebel}
and reviewed in the Appendix): in this sense his stability proof is
not fully convincing.  Mashhoon's analysis is based on perturbations
of test null rays in the unstable circular orbit of a KN black hole,
and is strictly valid only in the eikonal limit $l\gg 1$. Here we show
by a direct calculation that Mashhoon's claim is correct, and that his
predictions are surprisingly accurate even for small values of $l$. At
least for {\it scalar} perturbations our investigation of the QNM
spectrum can be considered conclusive, because for $s=0$ the DF
equation is {\it exact}.

One of us (KK) used the DF equation to compute the fundamental {\it
gravitational} QNM using WKB methods \cite{KNC}. The main problem of
this approach is not computational, but physical. The WKB
approximation is reasonably accurate for all values of $a$ and $Q$ (as
we show comparing results to a continued-fraction calculation in
Table~\ref{tab:l2m0}). However, the DF equation is derived under a
number of mathematical assumptions, and it is only {\it approximately}
valid for gravitational and EM perturbations of KN black holes.

A purpose of this paper is to clarify the physical range of validity
of the DF equation and the physical meaning of their
approximations. We first notice that the $a\to 0$ limit of the DF
equation does {\it not} yield any of the two Schr\"odinger-like
equations describing coupled EM-gravitational perturbations of the RN
black hole. Then we consider metric perturbations of the RN black hole
freezing EM perturbations (or, vice versa, EM perturbations freezing
the metric), and we compute the associated QNMs. We show that for
values of the charge $Q\lesssim M/2$ the results are in good (but {\it
not exact}) agreement with the DF equation. In other words, the
assumptions behind the separability conditions leading to the DF
equation are {\it not} equivalent to simply freezing EM (or
gravitational) perturbations. Nonetheless, the DF equation yields EM
and gravitational QNMs in good quantitative agreement with results for
the coupled EM-gravitational RN perturbation equations when $Q\lesssim
M/2$. Furthermore, the qualitative behavior of gravitational and EM
QNMs as functions of charge and angular momentum is very similar to
the results of our ``exact'' calculations for scalar
perturbations. Our calculations suggest that the DF equation provides
a reasonable approximation of the EM and gravitational dynamics of a
KN black hole, at least for the gravitationally small values of $Q$
expected in astrophysical applications.

The plan of the paper is as follows. In Sec.~\ref{sec:DFE} we present
the DF equation and we compute the corresponding QNM frequencies using
Leaver's continued fraction technique \cite{L}. Our calculation is
``exact'' (in the sense that the perturbation equations involve no
approximations) for $s=0$: as far as we know, it represents the first
example of numerical evidence for the stability of the KN metric to
scalar perturbations.  For EM and gravitational perturbations the RN
($a\to 0$) limit of the DF equation does not yield the standard RN
perturbation equations, which are summarized for completeness in
Sec.~\ref{sec:RN}. To clarify the meaning (and the limits of
applicability) of the DF equation for EM and gravitational
perturbations, in Sec.~\ref{sec:RNF} we consider approximate versions
of the RN perturbation equations. To our knowledge, the ``frozen''
picture we present in this Section has never been studied before. To
conclude we summarize our findings, discuss their meaning for concrete
applications in astrophysics and elementary particle models, and list
some open problems for future research. In the Appendix, for
completeness, we summarize results of Ref.~\cite{Mash} on perturbed
equatorial circular orbits of null rays in the KN metric and their
relation with the QNMs in the eikonal limit.


\section{Dudley-Finley perturbation equations for Kerr-Newman black holes}
\label{sec:DFE}

Dudley and Finley reduced the KN perturbation problem to a pair of
differential equations: one for the angular part of the perturbations,
and the other for the radial part \cite{DF,KNC}. In this paper we
consistently adopt Leaver's conventions \cite{L}. In particular, we
use geometrical units and set $2M=1$.  In Boyer-Lindquist coordinates,
defining $u=\cos\theta$, the angular equation
\be
\f{d}{du}\left[
(1-u^2)\f{d S_{lm}}{du}
\right]
+\left[
(a\omega u)^2-2a\omega su+s+ A_{lm}-{(m+su)^2\over 1-u^2}
\right]
S_{lm}=0\,,
\label{angularwaveeq}
\ee
is the same as in the Kerr case. The angular separation constant $E$
used in \cite{KNC} is related to Leaver's $A_{lm}$ by
$E=A_{lm}+s(s+1)$. Correcting a typo in Eq.~(5) of \cite{KNC} (the
factor $4\ii sK'$ should read $2\ii sK'$), the DF radial equation reads
\be\label{DFeq}
\Delta^{-s}\frac{d}{dr}
\left[\Delta^{s+1}\frac{dR}{dr}\right]
+\frac{1}{\Delta}\left[K^2-\ii s\frac{d\Delta}{dr} K+
\Delta\left(2\ii s \frac{dK}{dr}-\lambda\right)\right]R=0\,,
\ee
where $K\equiv (r^2+a^2)\omega-am$, $\Delta\equiv r^2-r+a^2+Q^2$ and
$\lambda=A_{lm}+(a\omega)^2-2am\omega$. The parameter $s=0,-1,-2$ for
scalar, EM and gravitational perturbations respectively, and $a$ is
the Kerr rotation parameter ($0\leq a<1/2$). In the Schwarzschild
limit $a\to 0$ the angular separation constant $A_{lm}\to
l(l+1)-s(s+1)$. The DF equation was also studied by Detweiler and Ove
\cite{DO}. It is exact only when the spin $s=0$ or when we set the
charge $Q=0$ (and then it reduces to the standard Teukolsky equation
for Kerr black holes \cite{teu}).

Let us consider the boundary conditions of Eq.~(\ref{DFeq}) at the
horizon. The horizon radius $r_+=(1+b)/2$, where $b\equiv
\sqrt{1-4(a^2+Q^2)}$. From the indicial equation it follows that the
ingoing solution at the horizon is such that
\be
R_{lm}\sim (r-r_+)^{-s-\ii\sigma_+} \mbox{~~as~~}r\to r_+\,,
\ee
where $\sigma_+=[\omega (r_+-Q^2)-am]/b$. Similarly, imposing purely
outgoing radiation at infinity we find
\be
R_{lm}\sim r^{-2s-1+\ii\omega} e^{\ii\omega r} \mbox{~~as~~}r\to \infty\,.  
\ee
Both these boundary conditions on the radial equation (\ref{DFeq}) and
the corresponding regularity conditions on the angular equation can be
cast as three-term continued fraction relations of the form
\be\label{CFKN}
0=\beta_0-
{\alpha_0\gamma_1\over \beta_1-}
{\alpha_1\gamma_2\over \beta_2-}\dots\,.
\ee
The coefficients of the radial continued fraction are:
\bea
\alpha_j&=&
j^2+
(2-2\ii \sigma_+-s)j
\\
&+&
\{
-32(\omega a^3m+\ii b^2\sigma_+)
+16
[\ii b^2s\sigma_+ +\ii samb+\omega^2a^4+a^2m^2+a^2\omega^2b
+b^2-b^2\sigma_+^2-am\omega b-\ii s\omega a^2b-sb^2]
\nn\\
&+&8
[b^2a^2\omega^2+a^2\omega^2
-am\omega-\ii b^2s\omega-b^2am\omega]
+4[\omega^2b+\omega^2b^3
-\ii b^3s\omega-\ii s\omega b]
+\omega^2(b^4+6b^2+1)
\}/(16b^2)\,,
\nn\\
\beta_j&=&-2j^2+
2[\ii \omega(b+1)+2\ii \sigma_+ -1]j
\nn\\
&+&
\{32\omega a^3m
+16[\ii b^2\sigma_+ +\omega b^3\sigma_++b^2\sigma_+ \omega
+b^2\sigma_+^2-a^2m^2-\omega^2a^4]
\nn\\
&+&8[b^2am\omega+\ii \omega b^3+\omega^2b^3+\ii b^2\omega
+am\omega-sb^2-a^2\omega^2-b^2-b^2A_{lm}]
+\omega^2(3b^4+6b^2-1)
\}/(8b^2)\,,
\nn\\
\gamma_j&=&j^2-
[2\ii (\omega+\sigma_+)-s]j
\nn\\
&-&
\{
32[\omega a^3m+b^2\sigma_+ \omega]
+24\ii b^2s\omega
+16[b^2\sigma_+^2+\ii b^2s\sigma_+
+a^2\omega^2b+\ii samb-am\omega b-\ii s\omega a^2b
-\omega^2a^4-a^2m^2]
\nn\\
&+&10\omega^2b^2
+8[b^2am\omega+am\omega-b^2a^2\omega^2-a^2\omega^2]
+4[\omega^2b^3+\omega^2b-\ii b^3s\omega-\ii s\omega b]
-\omega^2(1+b^4)
\}/(16b^2)\,.
\nn
\eea
and those of the angular continued fraction can be found in \cite{L}.
To find QNM frequencies we first fix the values of $a,~\ell,~m$ and
$\omega$, and find the angular separation constant $A_{lm}(\omega)$
looking for zeros of the {\it angular} continued fraction. Then we use
the corresponding eigenvalue to look for zeros of the {\it radial}
continued fraction as a function of $\omega$. The $n$--th quasinormal
frequency is (numerically) the most stable root of the $n$--th
inversion of the continued-fraction relation (\ref{CFKN}), i.e., it is
the root of
\bea\label{CFIKN}
&&\beta_n-
{\alpha_{n-1}\gamma_{n}\over \beta_{n-1}-}
{\alpha_{n-2}\gamma_{n-1}\over \beta_{n-2}-}\dots
{\alpha_{0}\gamma_{1}\over \beta_{0}}
={\alpha_n\gamma_{n+1}\over \beta_{n+1}-}
{\alpha_{n+1}\gamma_{n+2}\over \beta_{n+2}-}\dots\,,
\qquad (n=1,2,\dots)\,.
\eea
The infinite continued fraction appearing in equation (\ref{CFIKN})
can be summed ``bottom to top'' starting from some large truncation
index $N$. Nollert \cite{N} has shown that the convergence of the
procedure improves if such a sum is started using a series expansion
for the ``rest'' of the continued fraction, $R_N$, defined by the
equation
\be
R_N={\gamma_{N+1}\over\beta_{N+1}-\alpha_{N+1}R_{N+1}}\,.
\ee
The series expansion reads
\be\label{RN}
R_N=\sum_{k=0}^{\infty}C_k N^{-k/2}\,,
\ee
where the first few coefficients have the same form as in the Kerr
case \cite{BK}, except for the charge-dependent correction in $b$:
$C_0=-1$, $C_1=\pm\sqrt{-2\ii\omega b}$, $C_2=\left[ 3/4+\ii\omega
(b+1)-s \right]$.

\begin{table}
\centering
\caption{Fundamental scalar QNM of the KN metric with $l=m=0$ for
selected values of the charge and angular momentum.}
\vskip 12pt
\begin{tabular}{@{}cccccc@{}}
\hline
\hline
$a$    &$Q=0$                &$Q=0.1$              &$Q=0.2$              &$Q=0.3$              &$Q=0.4$ \\
\hline
\hline
0      &(0.220910,-0.209791) &(0.222479,-0.210122) &(0.227471,-0.210998) &(0.236910,-0.211849) &(0.253168,-0.210094)\\
0.1    &(0.221535,-0.209025) &(0.223121,-0.209319) &(0.228169,-0.210067) &(0.237712,-0.210596) &(0.254019,-0.207839)\\
0.2    &(0.223398,-0.206506) &(0.225033,-0.206673) &(0.230230,-0.206961) &(0.239994,-0.206302) &(0.255542,-0.199555)\\
0.3    &(0.226342,-0.201397) &(0.228021,-0.201267) &(0.233297,-0.200439) &(0.242568,-0.196687) &-\\
0.4    &(0.229074,-0.191402) &(0.230486,-0.190541) &(0.234035,-0.186969) &- &-\\
\hline
\hline
\end{tabular}
\label{tab:l0m0}
\end{table}

\begin{figure*}
\begin{center}
\begin{tabular}{cc}
\epsfig{file=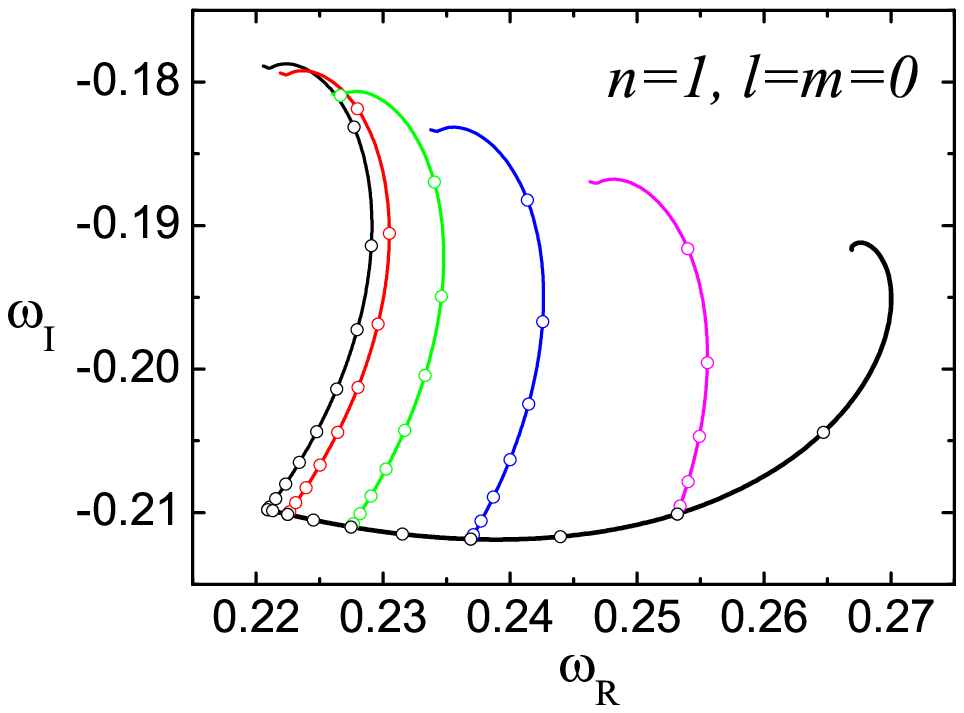,width=8.0cm,angle=0}&
\epsfig{file=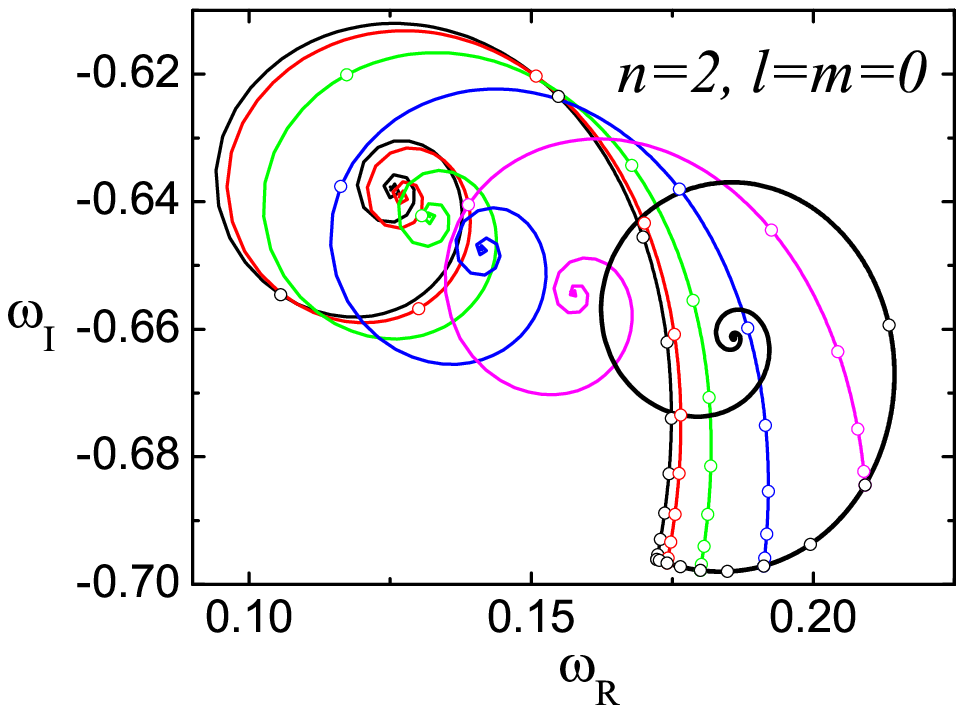,width=8.0cm,angle=0}\\
\epsfig{file=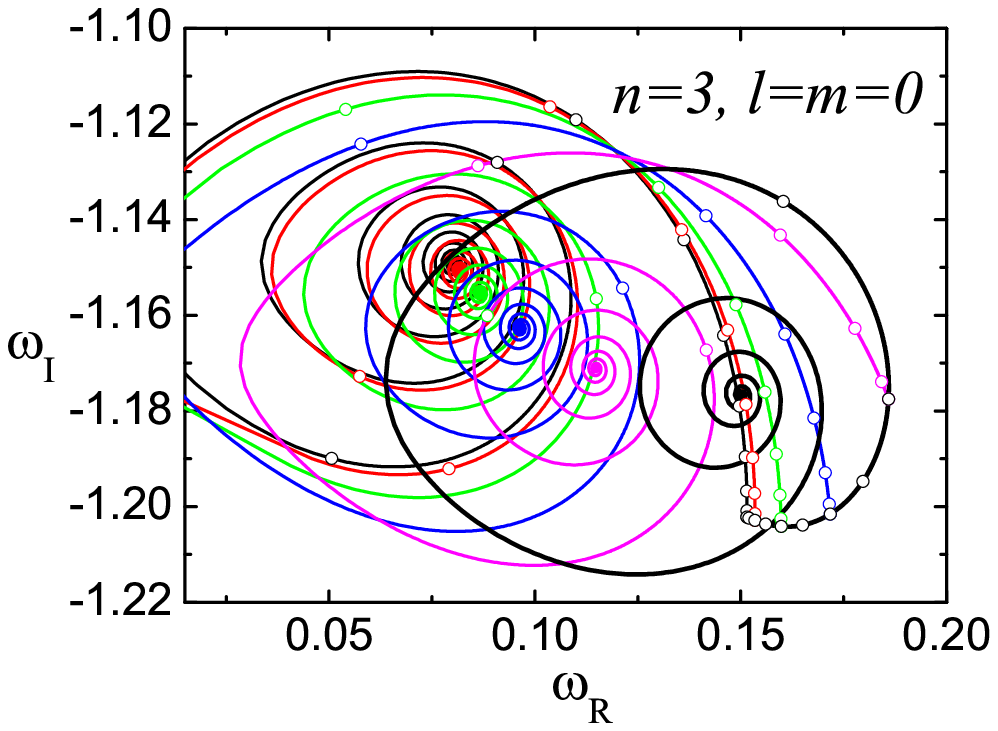,width=8.0cm,angle=0}&
\epsfig{file=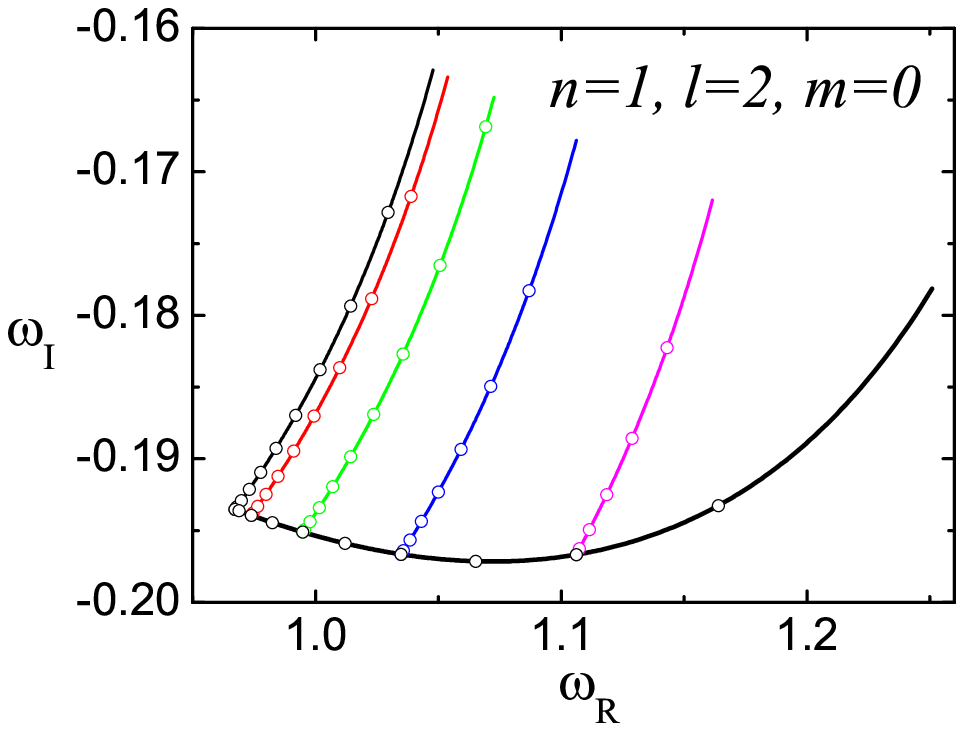,width=8.0cm,angle=0}\\
\end{tabular}
\caption{Left to right and top to bottom, the first three panels show
trajectories in the complex plane of the first three scalar QNMs of
the KN black hole for $l=m=0$. The thick black curve corresponds to
$a=0$ (RN limit). The thin curves are obtained increasing $a$ from
zero to the extremal limit for {\it fixed} values of the charge
($Q=0,~0.1,~0.2,~0.3,~0.4$ respectively). Open circles (when present)
mark selected values of the charge ($Q=0,~0.05,~0.1\dots$ along the
thick RN line) and angular momentum ($a=0,~0.05,~0.1\dots$ along the
thin lines). Bottom right panel: same plot for the fundamental mode
with $s=0$ and $l=2$. Thin lines are now trajectories of the $m=0$ KN
modes for $Q=0,~0.1,~0.2,~0.3,~0.4$. For clarity we do not show modes
with other values of $m$.
\label{fig:l0m0}
}
\end{center}
\end{figure*}

In Table~\ref{tab:l0m0} we give numerical results for the fundamental
scalar QNM with $l=m=0$ as a function of charge and angular
momentum. Dashed entries in this and the following Tables correspond
to combinations of $Q$ and $a$ for which $Q^2+a^2\geq M^2$.

Fig.~\ref{fig:l0m0} shows trajectories described by selected scalar
QNMs in the complex plane. In each panel, the thick black line
corresponds to modes of a RN (possibly charged but non-rotating) black
hole. As $Q$ increases the RN mode moves counterclockwise in the
complex frequency plane; open circles on the thick line mark
increasing values of the charge ($Q=0,~0.05,~0.1,\dots$). For five
fixed values of $Q$ (namely $Q=0,~0.1,~0.2,~0.3,~0.4$) we plot the KN
QNM trajectories as we increase $a$: the results are the five thin
lines branching from the RN limit. Open circles on these curves mark
increasing values of the angular momentum
($a=0,~0.05,~0.1,\dots$). The observed spiraling behavior (as a
function of both charge $Q$ and angular momentum $a$) confirms and
extends results presented in \cite{KS,O,AO,BK,BCKO,BCY,GA} and
summarized in \cite{milano}. The bottom-right panel shows trajectories
of the fundamental scalar mode with $l=2$, $m=0$. As we increase $l$
the counterclockwise bending (that for $l=0$ is very pronounced even
for the lowest overtones) only shows up at higher overtone indices
($n\sim 10$ for $l=2$).

\begin{figure*}
\begin{center}
\begin{tabular}{ccc}
\epsfig{file=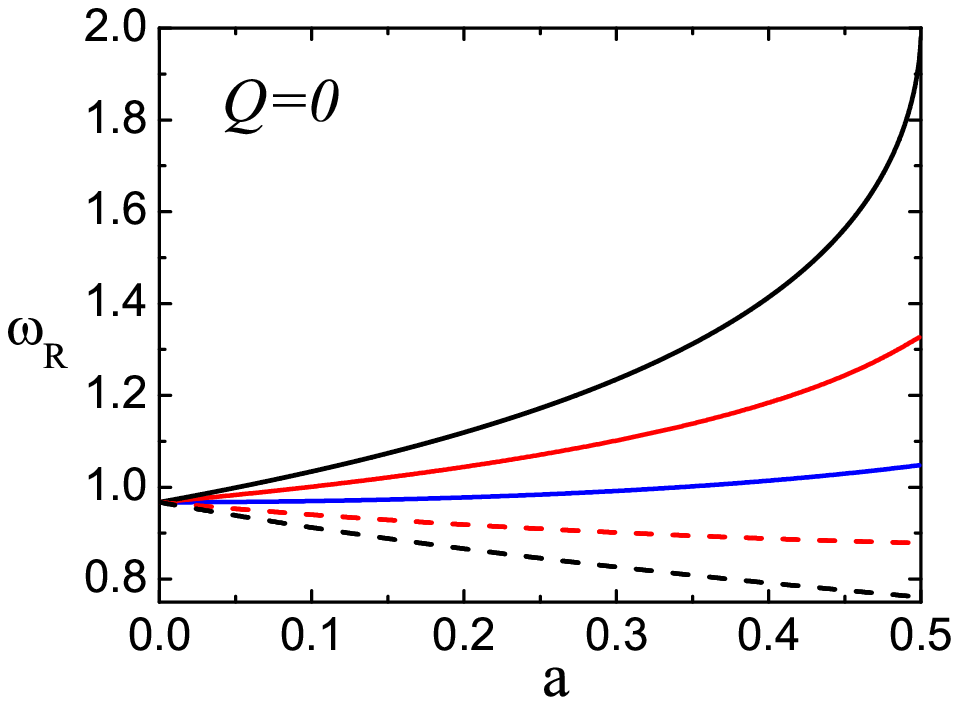,width=6.0cm,angle=0}&
\epsfig{file=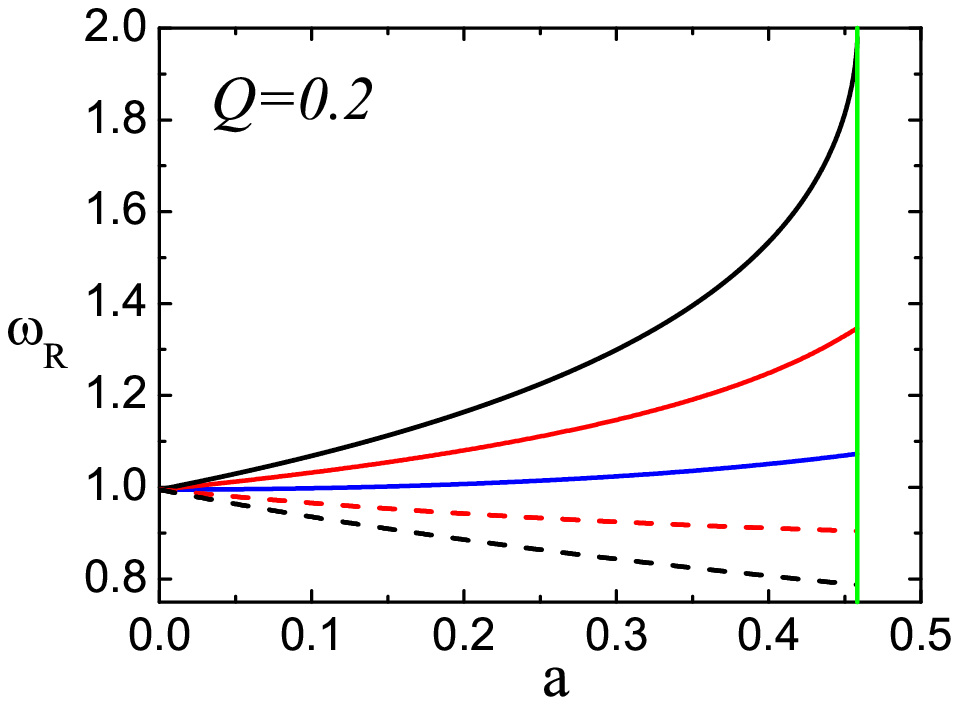,width=6.0cm,angle=0}&
\epsfig{file=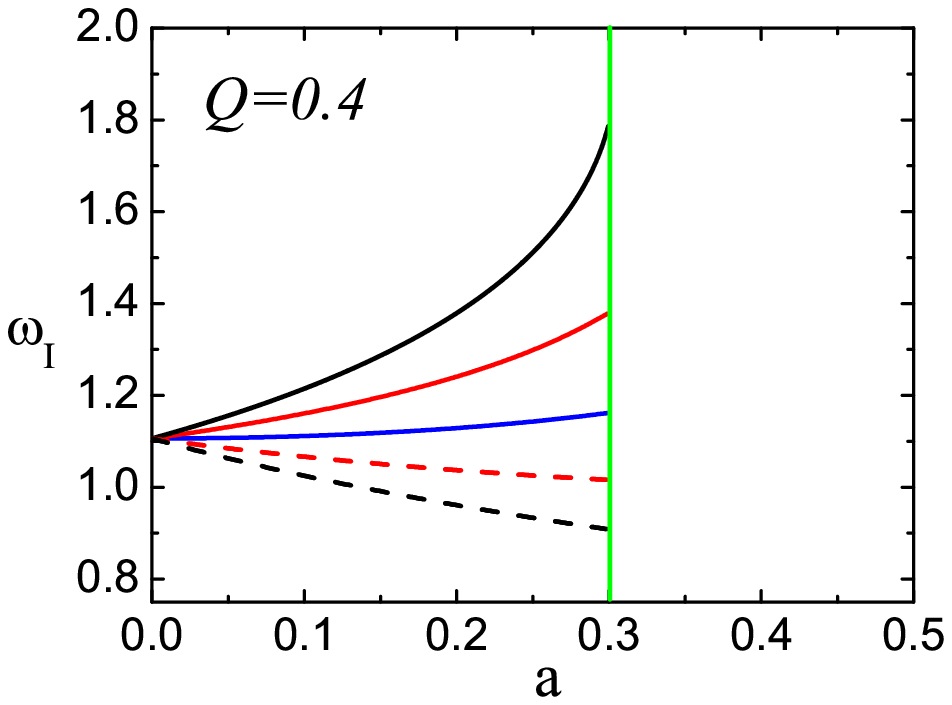,width=6.0cm,angle=0}\\
\end{tabular}
\caption{Real part of the fundamental scalar QNM of a KN black hole
for $l=2$ and different values of $m$. Curves from top to bottom refer
to $m=2,1,0,-1,-2$. The three panels correspond to different values of
$Q$, as indicated. For $Q\neq 0$, the vertical line marks the extremal
limit.
\label{fig:l2r}
}
\end{center}
\end{figure*}

\begin{figure*}
\begin{center}
\begin{tabular}{ccc}
\epsfig{file=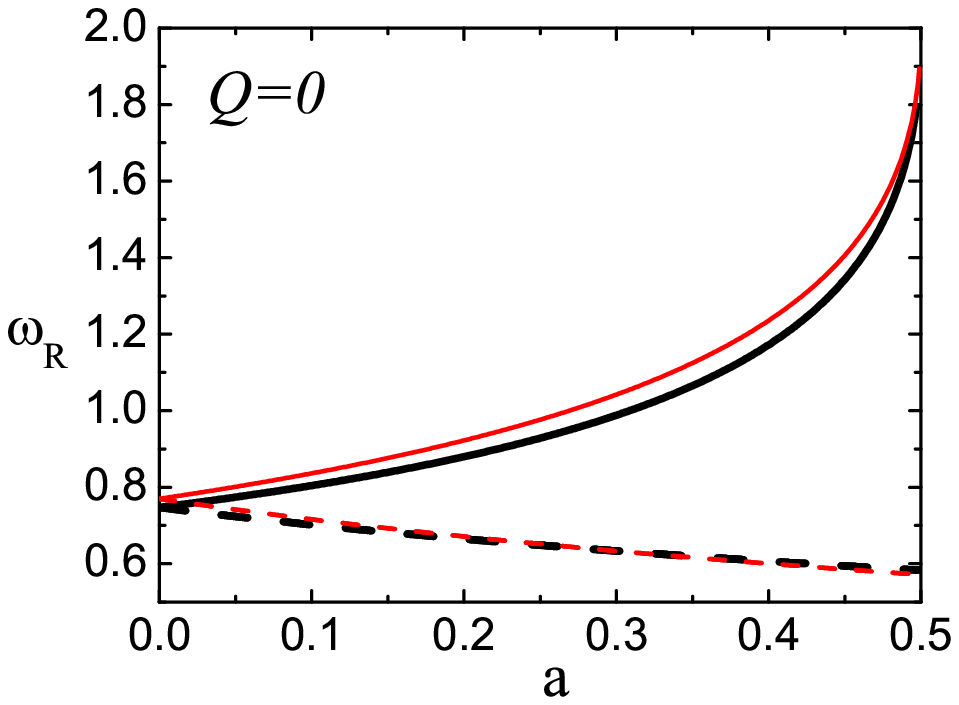,width=6.0cm,angle=0}&
\epsfig{file=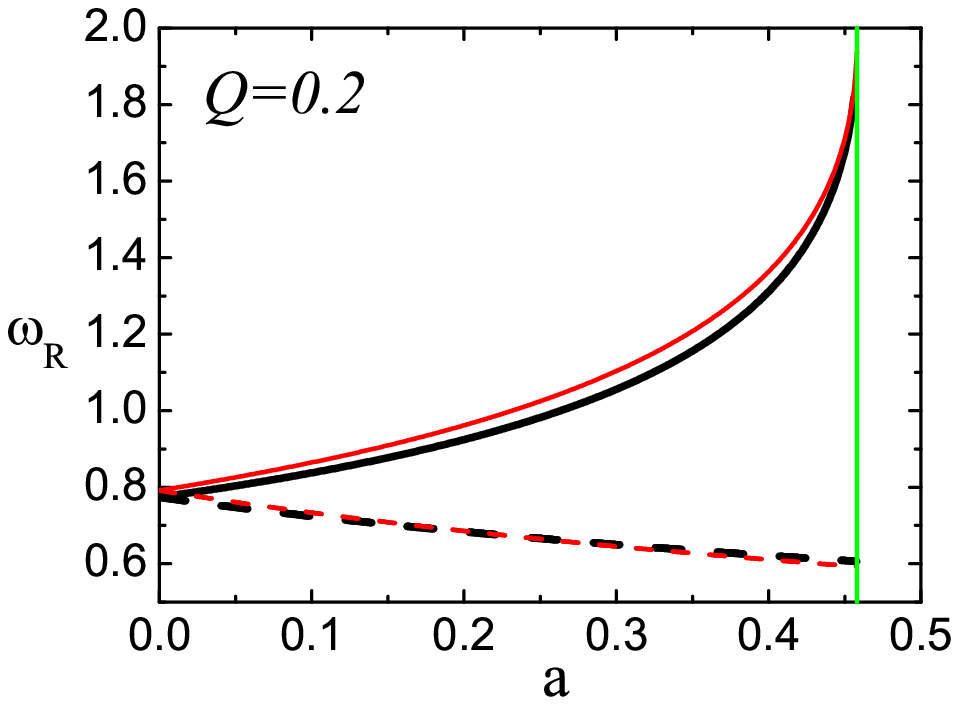,width=6.0cm,angle=0}&
\epsfig{file=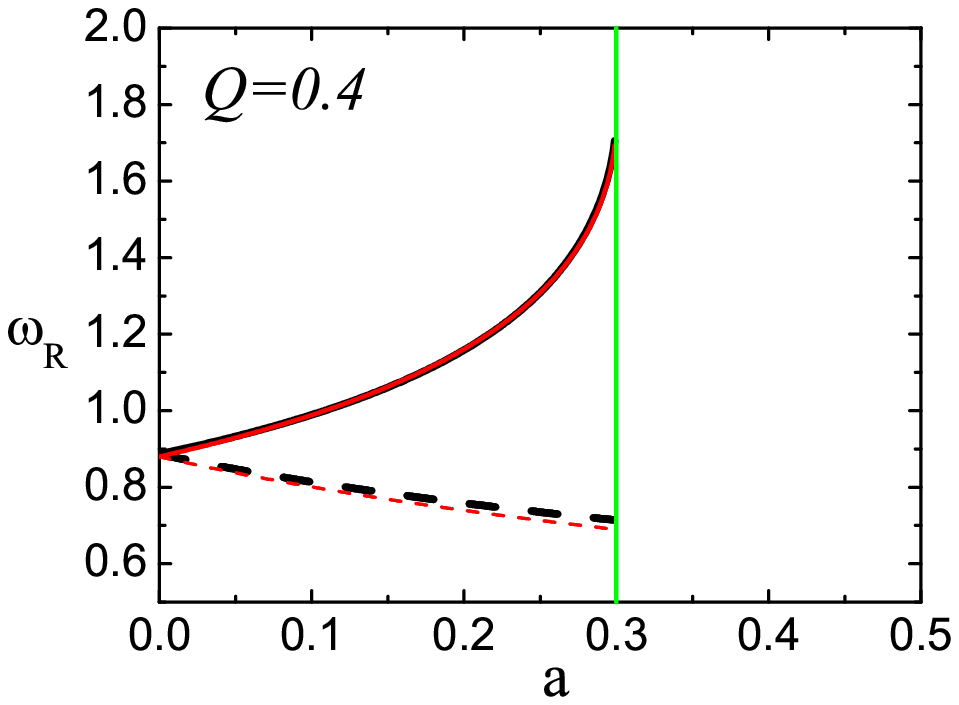,width=6.0cm,angle=0}\\
\epsfig{file=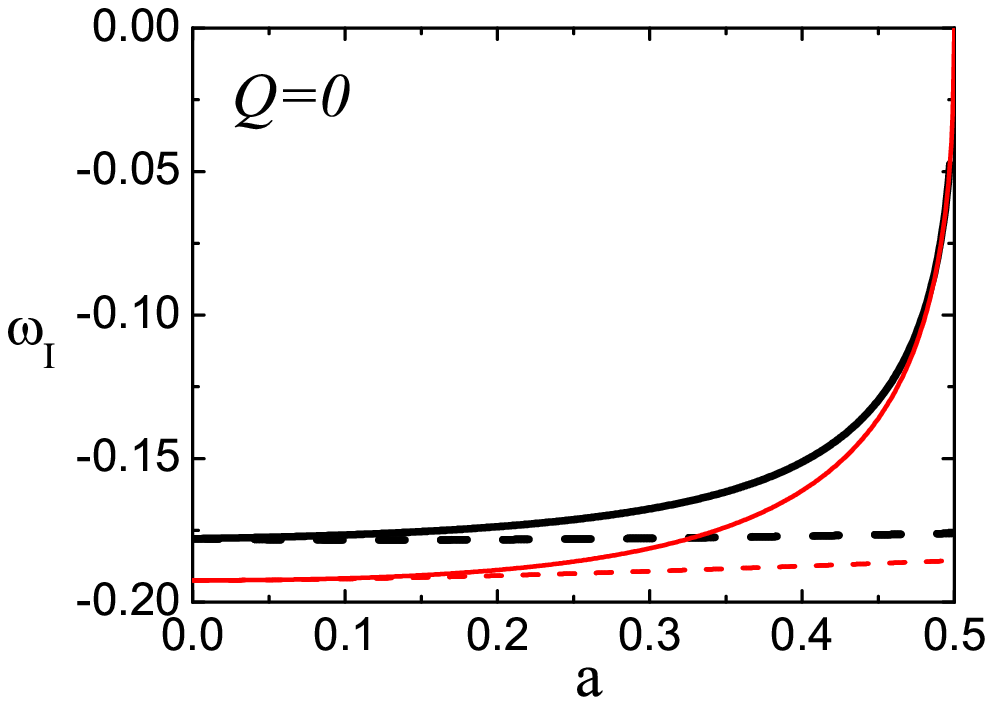,width=6.0cm,angle=0}&
\epsfig{file=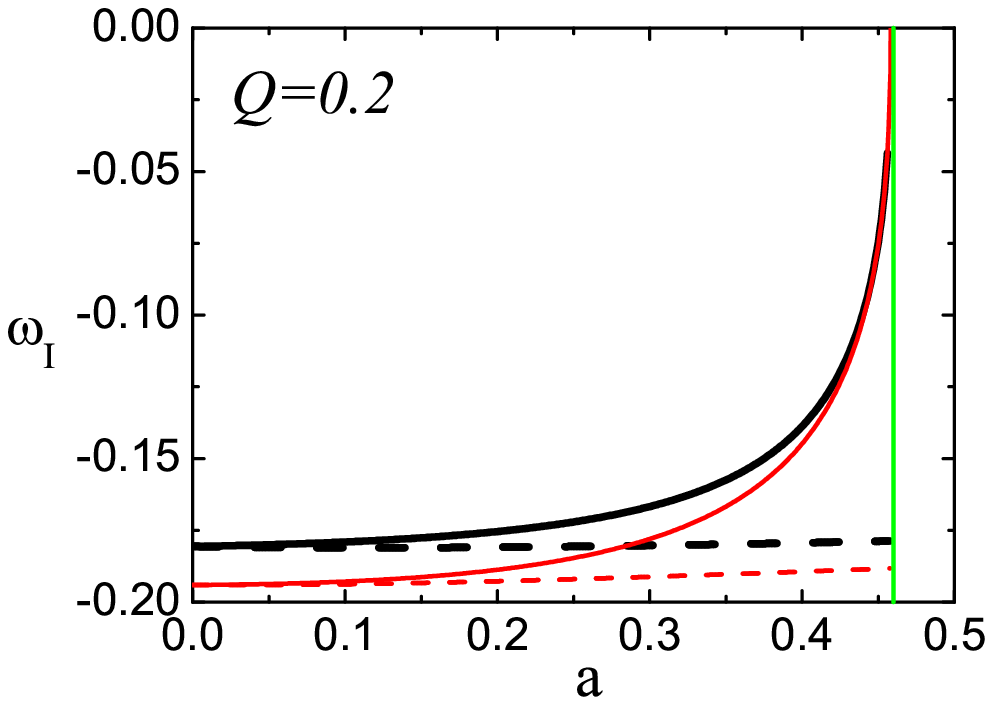,width=6.0cm,angle=0}&
\epsfig{file=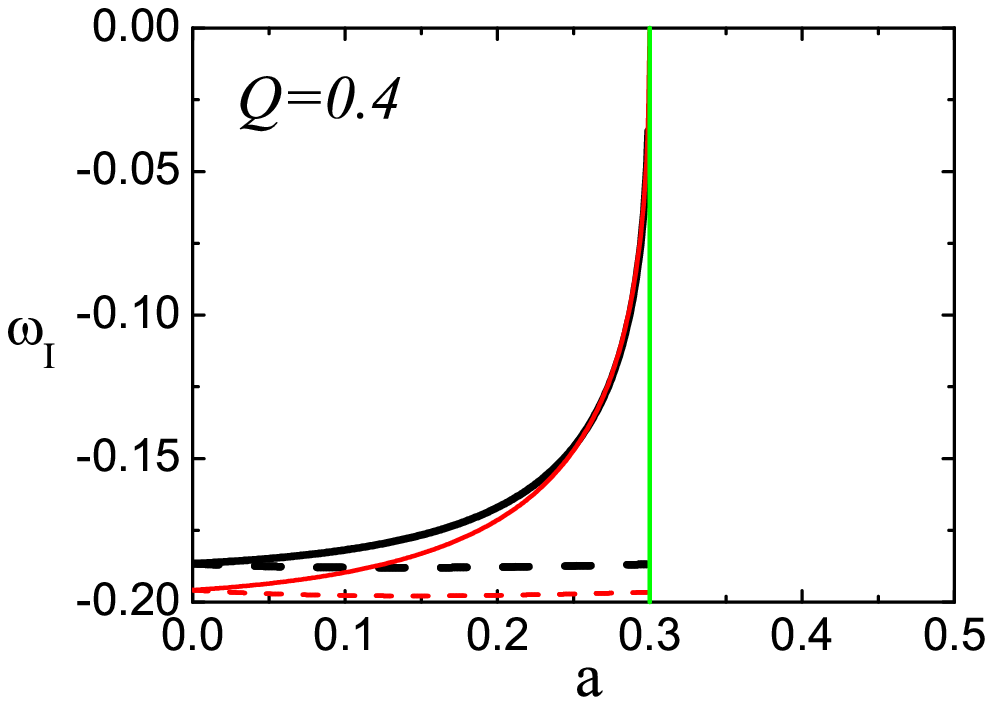,width=6.0cm,angle=0}\\
\end{tabular}
\caption{Real part (top) and imaginary part (bottom) of the
fundamental $l=2$ gravitational QNM of a KN black hole. Solid lines
refer to $m=2$, dashed lines to $m=-2$. Different panels correspond to
different values of the charge, as indicated. For $Q\neq 0$, the
vertical line marks the extremal limit. Thick (black) lines are
numerical results from the DF equation. Thin (red) lines are obtained
from Mashhoon's arguments, which are strictly valid only as $l\to
\infty$ (see Appendix). Considering that we are looking at modes with
$l=2$ the agreement of Mashhoon's predictions with numerical results
is quite impressive, particularly for the real part of the fundamental
QNM frequency and for large values of $Q$ and $a$.
\label{fig:mash}
}
\end{center}
\end{figure*}

When $l>0$, and for any value of the charge $0\leq Q<1/2$, rotation
induces a Zeeman-like splitting of the modes. In Fig.~\ref{fig:l2r} we
show the splitting of the real part of the frequency of the
fundamental scalar QNM with $l=2$ for three representative values of
the charge (from left to right: $Q=0,~0.2,~0.4$). For Kerr black
holes, the lowest-lying modes with $l=m$ tend to approach the critical
frequency for superradiance $m\Omega$ (where $\Omega$ is the
rotational velocity of the black hole horizon) in the extremal limit.
Detweiler \cite{De} first presented analytical arguments to explain
this clustering of modes at the superradiant frequency in the extremal
limit. The implicit assumptions in his argument have recently been
re-examined in \cite{GADet,CDet}. In the general KN case the extremal
frequencies for modes with $l=|m|$ have analytically been computed by
Mashhoon \cite{Mash}. We quote his result, which is in excellent
agreement with our numerical calculations (see Fig.~\ref{fig:mash}
below), at the end of the Appendix.

In Table~\ref{tab:l2m0} we consider gravitational perturbations with
$l=2$, $m=0$. To assess the reliability of WKB methods in the KN case
we compare results from Leaver's continued fraction approach (which
can be considered exact, within the given numerical accuracy) with the
third order WKB technique used in \cite{KNC}, building on previous
results in \cite{sai,KerrQNM}. The agreement for the fundamental
gravitational QNM with $l=2$, $m=0$ is excellent, typical errors being
smaller than one part in a thousand. The WKB approach systematically
underestimates the real parts and overestimates the imaginary
parts. Third order WKB methods become slightly less reliable for large
rotation parameter $a$; at fixed $a$, their accuracy depends very
weakly on $Q$.  Table~\ref{tab:l2m2} shows results for the $l=m=2$
``barlike'' mode, which is generally believed to be dominant in the
quasinormal ringing following gravitational collapse to a rotating
black hole \cite{FHH}.

\begin{table}
\centering
\caption{Comparison of QNM frequencies of the DF equation obtained by
the continued fraction method (first row for each value of $a$) with
third-order WKB results from \cite{KNC} (second row). Numbers refer to
the fundamental gravitational QNM with $l=2$, $m=0$.}
\vskip 12pt
\begin{tabular}{@{}cccccc@{}}
\hline
\hline
$a$    & $Q=0$                & $Q=0.1$              & $Q=0.2$              & $Q=0.3$              & $Q=0.4$ \\
\hline
\hline
0      & (0.747343,-0.177925) & (0.753643,-0.178602) & (0.773902,-0.180620) & (0.813278,-0.183793) & (0.886468,-0.186603) \\
       & (0.7463,-0.1784)     & (0.7530,-0.1792)     & (0.7734,-0.1812)     & (0.8128,-0.1844)     & (0.8860,-0.1870) \\
\hline
0.1    & (0.750248,-0.177401) & (0.756664,-0.178053) & (0.777325,-0.179977) & (0.817630,-0.182905) & (0.893230,-0.184893) \\
       & (0.7496,-0.1780)     & (0.7560,-0.1786)     &(0.7835,-0.1804)& (0.8172,-0.1834)     & (0.8928,-0.1854) \\
\hline
0.2    & (0.759363,-0.175653) & (0.766159,-0.176211) & (0.788148,-0.177790) & (0.831579,-0.179789) & (0.915681,-0.178292) \\
       & (0.7586,-0.1762)     & (0.7654,-0.1766)     & (0.7874,-0.1782)     & (0.8310,-0.1802)     & (0.9150,-0.1788) \\
\hline
0.275  & (0.771072,-0.173158) & (0.778394,-0.173563) & (0.802246,-0.174561) & (0.850222,-0.174875) & (0.947708,-0.165369) \\
       & (0.7702,-0.1736)     & (0.7776,-0.1740)     & (0.8014,-0.1750)     & (0.8494,-0.1754)     & (0.9470,-0.1660) \\
\hline
0.3    & (0.776108,-0.171989) & (0.783669,-0.172315) & (0.808382,-0.173003) & (0.858524,-0.172355) & - \\
       & (0.7752,-0.1724)     & (0.7828,-0.1728)     & (0.8074,-0.1734)     & (0.8578,-0.1728)     & - \\
\hline
0.4    & (0.803835,-0.164313) & (0.812886,-0.163977) & (0.843068,-0.161921) & -                    & - \\
       & (0.8026,-0.1648)     & (0.8118,-0.1646)     & (0.8410,-0.1626)     & -                    & - \\
\hline
0.45   & (0.824009,-0.156965) & (0.834319,-0.155756) & (0.869094,-0.149647) & -                    & - \\
       & (0.8228,-0.1576)     & (0.8330,-0.1564)     & (0.8680,-0.1506)     & -                    & - \\
\hline
0.48   & (0.838981,-0.150055) & (0.850244,-0.147806) & -                    & -                    & - \\
       & (0.8378,-0.1510)     & (0.8490,-0.1488)     & -                    & -                    & - \\
\hline
\hline
\end{tabular}
\label{tab:l2m0}
\end{table}

\begin{table}
\centering
\caption{QNM frequencies of the DF equation obtained by the continued
fraction method. Numbers refer to the fundamental gravitational QNM
with $l=m=2$.}
\vskip 12pt
\begin{tabular}{@{}cccccc@{}}
\hline
\hline
$a$    & $Q=0$                & $Q=0.1$              & $Q=0.2$              & $Q=0.3$              & $Q=0.4$ \\
\hline
\hline

0      & (0.747343,-0.177925) & (0.753643,-0.178602) & (0.773902,-0.180620) & (0.813278,-0.183793) & (0.886468,-0.186603) \\
0.1    & (0.804291,-0.176622) & (0.812131,-0.177253) & (0.837615,-0.179080) & (0.888479,-0.181653) & (0.989234,-0.181863) \\
0.2    & (0.879684,-0.173764) & (0.890062,-0.174238) & (0.924394,-0.175451) & (0.996178,-0.176132) & (1.158394,-0.167066) \\
0.275  & (0.956470,-0.169604) & (0.970151,-0.169774) & (1.016478,-0.169756) & (1.120103,-0.166278) & (1.432934,-0.122011) \\
0.3    & (0.988090,-0.167530) & (1.003384,-0.167519) & (1.055787,-0.166749) & (1.177413,-0.160442) & - \\
0.4    & (1.172034,-0.151259) & (1.200883,-0.149228) & (1.311304,-0.138796) & -                    & - \\
0.45   & (1.343229,-0.129738) & (1.395907,-0.123109) & (1.678651,-0.074535) & -                    & - \\
0.48   & (1.535348,-0.098867) & (1.653330,-0.077666) & -                    & -                    & - \\
\hline
\hline
\end{tabular}
\label{tab:l2m2}
\end{table}

Perhaps the most important outcome of our analysis is a null result:
{\it we could not find any unstable mode}. Our quasinormal mode
calculation confirms qualitatively Mashhoon's arguments in favor of
the stability of the KN metric \cite{Mash}. Fig.~\ref{fig:mash} shows
that the agreement is more than qualitative. In fact his argument
(which is based on the eikonal approximation, and is expected to be
accurate only for $l\gg 1$) captures almost perfectly the behavior of
the fundamental gravitational QNMs with $l=|m|=2$ as functions of
charge $Q$ and angular momentum $a$. This agreement is particularly
good for the real part of the frequencies, and it can be exploited for
``quick and dirty'' estimates of QNM excitation in astrophysical
scenarios (see e.g. \cite{AG}).

A major problem of the DF equation is that it only provides an {\it
approximation} to the problem of coupled EM-gravitational
perturbations of the KN black hole. One can easily check that the
limit $Q\to 0$ of Eq.~(\ref{DFeq}) yields the Teukolsky equation
\cite{teu} and the corresponding QNM frequencies. In fact, the first
column of our Table~\ref{tab:l2m0} is in perfect agreement with
Table~2 in \cite{L}. However, as $a\to 0$ (first line of
Table~\ref{tab:l2m0}) we do not recover the QNM frequencies of the RN
metric. This is no surprise: for the reasons discussed in the
Introduction we expect the DF equation to be a good approximation only
for small values of $Q$. Below we will show that this is indeed the
case, and give a quantitative meaning to the above statement.

\section{Reissner-Nordstr\"om: the standard treatment}
\label{sec:RN}

Here we briefly summarize the computational procedure for RN black
holes ($a=0$). More details can be found in Refs.~\cite{L,N} and
especially \cite{L2}. This Section is primarily intended to establish
notation for the ``frozen'' approximation to be introduced in Section
\ref{sec:RNF}.

It is well known that polar perturbations of the RN black hole can be
obtained from the axial perturbations by a Chandrasekhar
transformation \cite{MTB}; in particular, the QNM spectra of polar and
axial perturbations are the same. For this reason in the following we
only consider axial perturbations, that we denote by a superscript
$(-)$. Let us introduce a tortoise coordinate $r_*$ by the usual
relation
\be\label{tortoise}
{d r\over d r_*}={\Delta\over r^2},
\ee
where $\Delta=r^2-r+Q^2$ (recall that in our units $2M=1$ and $0\leq
Q<1/2$). Explicitly, the tortoise coordinate can be written as
\be
r_*=r+\frac{r_+^2}{r_+-r_-}\ln(r-r_+)-\frac{r_-^2}{r_+-r_-}\ln(r-r_-)\,,
\ee 
where $r_\pm=(1\pm\sqrt{1-4Q^2})/2$ is the location of the inner
(Cauchy) and outer (event) horizons of the RN metric. Using tensor
spherical harmonics to separate the angular dependence and a Fourier
decomposition to get rid of the time dependence, EM and axial
gravitational perturbations of the RN metric are described by two
coupled wave equations:
\bea\label{axialcoupled}
\left({d^2\over dr_*^2}+\omega^2\right)H_2^{(-)}&=&
\frac{\Delta}{r^5}\left[l(l+1)r-\f{3}{2}+\frac{4Q^2}{r}\right]H_2^{(-)}
-\f{3}{2}H_2^{(-)}+2Q\sqrt{(l-1)(l+2)}H_1^{(-)}\,,\\
\left({d^2\over dr_*^2}+\omega^2\right)H_1^{(-)}&=&
\frac{\Delta}{r^5}\left[l(l+1)r-\f{3}{2}+\frac{4Q^2}{r}\right]H_1^{(-)}
+\f{3}{2}H_1^{(-)}+2Q\sqrt{(l-1)(l+2)}H_2^{(-)}\,, \nn
\eea
where $H_2^{(-)}$ corresponds to perturbations of the gravitational
field and $H_1^{(-)}$ to perturbations of the EM field
\cite{MTB,Burko}. The usual procedure is to decouple the system
(\ref{axialcoupled}) to obtain Schr\"odinger-like equations of the
form
\be\label{axialRN}
\left({d^2\over dr_*^2}+\omega^2\right)Z_i^{(-)}=V_i^{(-)}Z_i^{(-)}\,,
\ee
where
\be\label{RNpot}
V_i^{(-)}=\frac{\Delta}{r^5}\left[l(l+1)r-q_j+\frac{4Q^2}{r}\right]\,,\qquad
(i,j=1,2,~i\neq j)\,,
\ee
and
\be\label{q12}
q_1=\left[3+\sqrt{9+16Q^2(l-1)(l+2)}\right]/2\,,\qquad
q_2=\left[3-\sqrt{9+16Q^2(l-1)(l+2)}\right]/2\,.
\ee
The price to pay is that, unlike $H_1^{(-)}$ and $H_2^{(-)}$, the
decoupled radial functions $Z_1^{(-)}$ and $Z_2^{(-)}$ are not simply
degrees of freedom of the EM and gravitational field for any $Q\neq
0$.  Only in the limit $Q=0$ do the potentials $V_1^{(-)}$ and
$V_2^{(-)}$ describe, respectively, purely EM and axial--gravitational
perturbations of a Schwarzschild black hole. The radial equations
(\ref{axialRN}) are solved by a series expansion of the form
\be\label{RNser}
Z_i^{(-)}=\frac{r_+}{r}e^{-2\ii\omega r_+}(r_+-r_-)^{-2\ii\omega-1}
(r-r_-)^{1+\ii\omega}e^{\ii\omega r}u^{-\ii\omega r_+^2/(r_+-r_-)}
\sum_{j=0}^\infty a_j u^j\,,
\ee
where $u=(r-r_+)/(r-r_-)$ and the coefficients $a_j$ are determined by
a four-term recursion relation. The problem can be reduced to a
three-term recursion relation of the form (\ref{CFKN}) using a
Gaussian elimination step \cite{L2}. Then we can use standard
techniques to determine the quasinormal frequencies. The convergence
of the summation can be improved using Nollert's expansion (\ref{RN})
for the rest. The first few coefficients of Nollert's series can be
found in \cite{BK}.

\section{Freezing Reissner-Nordstr\"om: an approximate decoupling}
\label{sec:RNF}

If we are interested in the oscillations of an astrophysical black
hole we can usually assume the black hole charge (in geometrized
units) to be small, $Q\ll M$. In this limit it is reasonable to ignore
the coupling between the EM field and the metric. In the following we
study QNM frequencies of the RN black hole ignoring the mutual effect
of the EM field on the metric and vice versa. We simply decouple the
system (\ref{axialcoupled}) {\it setting to zero the EM
(gravitational) perturbations} $H_1^{(-)}$ ($H_2^{(-)}$). The
resulting equations are:
\be\label{sumeq}
\left[\frac{d^2}{dr_*^2}+\omega^2-V^*_i \right]H_i^{(-)}=0\,,\qquad (i=1,2)\,.
\ee
The gravitational potential when we freeze EM perturbations,
$H_1^{(-)}=0$, is
%
\be\label{decoup}
V^*_2=\frac{\Delta}{r^5}\left[l(l+1)r-3+\frac{4Q^2}{r}\right]\,.
\ee
$V_2^*$ reduces to the standard RW potential for $Q=0$. The EM
potential freezing metric perturbations, $H_2^{(-)}=0$, becomes
%
\be\label{decoupEM}
V^*_1=\frac{\Delta}{r^5}\left[l(l+1)r+\frac{4Q^2}{r}\right]\,.
\ee
Similarly to $V_2^*$, $V_1^*$ reduces to the potential for EM
perturbations of Schwarzschild black holes when $Q=0$.  The two
potentials can be written in the compact form
\be\label{dec}
V^*_i=\frac{\Delta}{r^5}\left[l(l+1)r+(1-s^2)+\frac{4Q^2}{r}\right]\,,\qquad
(i=|s|=1,2)\,.  
\ee
For completeness we also consider scalar perturbations of the RN black
hole: in this case no approximations are involved, and the potential
reads \cite{BKRN}
\be\label{sca}
V_0=\frac{\Delta}{r^5}\left[l(l+1)r+1-\frac{2Q^2}{r}\right]\,.
\ee

\begin{table}
\centering
\caption{Scalar, EM and gravitational QNMs of the RN black hole. The
first line gives third order WKB results, and the second line sixth
order WKB results for the decoupled potentials (\ref{dec}) and
(\ref{sca}). Sixth order WKB results have been kindly provided by
Vitor Cardoso. The third line gives Leaver's results for the DF
equation in the limit $a\to 0$. In the EM and gravitational cases,
the fourth line gives results from Leaver's method applied to the
coupled EM-gravitational potentials $V^{(-)}_i$ ($i=1,2$).}
\vskip 12pt
\begin{tabular}{@{}cccccc@{}}
\hline
\hline
Scalar, $l=0$& $Q=0$                & $Q=0.1$              & $Q=0.2$              & $Q=0.3$              & $Q=0.4$ \\
\hline
WKB3, $V_0$        &(0.209290,-0.230390) &(0.210984,-0.230440) &(0.216361,-0.230392) &(0.226237,-0.229299) &(0.240493,-0.223521)\\
WKB6, $V_0$        &(0.220939,-0.201628) &(0.222496,-0.202043) &(0.227572,-0.203102) &(0.237028,-0.204762) &(0.251727,-0.207292)\\
Leaver, DF         &(0.220910,-0.209791) &(0.222479,-0.210122) &(0.227471,-0.210998) &(0.236910,-0.211849) &(0.253168,-0.210094)\\
\hline
\hline
Scalar, $l=2$& $Q=0$                & $Q=0.1$              & $Q=0.2$              & $Q=0.3$              & $Q=0.4$ \\
\hline
WKB3, $V_0$        &(0.966422,-0.193610) &(0.973006,-0.194022) &(0.993994,-0.195186) &(1.033982,-0.196707) &(1.105332,-0.196689)\\
WKB6, $V_0$        &(0.967284,-0.193532) &(0.973859,-0.193948) &(0.994823,-0.195120) &(1.034772,-0.196664) &(1.106104,-0.196689)\\
Leaver, DF         &(0.967288,-0.193518) &(0.973863,-0.193933) &(0.994826,-0.195107) &(1.034775,-0.196649) &(1.106105,-0.196675)\\
\hline
\hline
EM, $l=1$& $Q=0$                & $Q=0.1$              & $Q=0.2$              & $Q=0.3$              & $Q=0.4$ \\
\hline
WKB3, $V^*_1$      &(0.491740,-0.186212) &(0.498236,-0.186976) &(0.519336,-0.189280) &(0.561324,-0.193009) &(0.643293,-0.196606)\\
WKB6, $V^*_1$      &(0.496383,-0.185274) &(0.502809,-0.186118) &(0.523726,-0.188648) &(0.565304,-0.192902) &(0.646616,-0.197253)\\
Leaver, DF         &(0.496527,-0.184975) &(0.500367,-0.185468) &(0.512675,-0.186892) &(0.536404,-0.188917) &(0.579729,-0.189708)\\
Leaver, $V^{(-)}_1$&(0.496527,-0.184975) &(0.502950,-0.185805) &(0.523843,-0.188312) &(0.565513,-0.192408) &(0.646987,-0.196545)\\
\hline
\hline
EM, $l=2$& $Q=0$                & $Q=0.1$              & $Q=0.2$              & $Q=0.3$              & $Q=0.4$ \\
\hline
WKB3, $V^*_1$      &(0.914262,-0.190130) &(0.922188,-0.190667) &(0.947700,-0.192226) &(0.997397,-0.194478) &(1.090249,-0.195485)\\
WKB6, $V^*_1$      &(0.915187,-0.190022) &(0.923104,-0.190565) &(0.948588,-0.192142) &(0.998235,-0.194426) &(1.091011,-0.195485)\\
Leaver, DF         &(0.915191,-0.190009) &(0.921657,-0.190451) &(0.942307,-0.191710) &(0.981812,-0.193421) &(1.052923,-0.193730)\\
Leaver, $V^{(-)}_1$&(0.915191,-0.190009) &(0.925930,-0.190747) &(0.959852,-0.192884) &(1.024022,-0.196033) &(1.140260,-0.198138)\\
\hline
\hline
Gravitational, $l=2$& $Q=0$                & $Q=0.1$              & $Q=0.2$              & $Q=0.3$              & $Q=0.4$ \\
\hline
WKB3, $V^*_2$      &(0.746324,-0.178435) &(0.751578,-0.178831) &(0.768355,-0.179955) &(0.800439,-0.181460) &(0.858116,-0.181608)\\ 
WKB6, $V^*_2$      &(0.747239,-0.177781) &(0.752520,-0.178204) &(0.769379,-0.179415) &(0.801591,-0.181106) &(0.859434,-0.181619)\\
Leaver, DF         &(0.747343,-0.177925) &(0.753643,-0.178602) & (0.773902,-0.180620) & (0.813278,-0.183793) & (0.886468,-0.186603) \\
Leaver, $V^{(-)}_2$&(0.747343,-0.177925) &(0.749489,-0.178150) &(0.756874,-0.178796) &(0.772435,-0.179627) &(0.802434,-0.179286)\\
\hline
\hline
\end{tabular}
\label{RNapprox}
\end{table}

In Table~\ref{RNapprox} we present the fundamental scalar, EM and
gravitational QNM frequency for five selected values of the
charge. From top to bottom we consider scalar perturbations with $l=0$
and $l=2$, EM perturbations with $l=1$ and $l=2$ and gravitational
perturbations with $l=2$.  

We computed QNMs of the potentials (\ref{dec}) and (\ref{sca}) using a
third order WKB expansion \cite{IW}, then we checked convergence using
a sixth order WKB expansion \cite{Kon}. Sixth order results have been
kindly provided by Vitor Cardoso. WKB results from the third (sixth)
order expansion are given in the rows marked WKB3 (WKB6). ``Leaver,
DF'' means that QNM frequencies have been computed using continued
fractions and the $a=0$ limit of the DF equation (\ref{DFeq}) with the
appropriate value of $s$.  Finally, ``Leaver, $V_i^{(-)}$'' means we
applied Leaver's continued fraction method to the coupled
EM-gravitational system (\ref{axialRN}): the results are the ``true''
oscillation frequencies of the RN black hole \cite{L2,BK}.
Some comments are in order: 

\begin{itemize}

\item[i)] The sixth order WKB results can be considered reliable. Even
in the case in which we expect the worst convergence (scalar
perturbations with $l=0$) a sixth order WKB expansion agrees very well
with results from the DF equation (which in this case, we stress it
again, is {\it exact}). For scalar modes with $l=2$ the agreement
between the sixth order WKB and the results from the DF equation is
quite astonishing. This is really a double check: not only it proves
that WKB results can be considered reliable, it also shows (by a
completely independent calculation) that the scalar QNMs obtained in
Sec.~\ref{sec:DFE} have the correct limit as $a\to 0$.

\item[ii)] The perfect agreement between decoupled and coupled EM
perturbations with $l=1$ is no surprise. Mathematically, a quick
inspection of Eq.~(\ref{axialRN}) with $i=1$ reveals that, since for
$l=1$ we have $q_2=0$ for any value of $Q$, EM perturbations with
$l=1$ are always decoupled from gravitational perturbations. Of
course, physically this decoupling is due to the nonradiative
character of dipolar gravitational fields (the first radiative
multipole being the quadrupole, $l=2$). Notice however that for EM
fields and $l=1$ QNMs of the DF equation disagree with the other two
approaches, the deviations increasing for large values of $Q$. This is
evidence that {\it when we consider the DF equation with $s=-1$ we are
not really ``killing'' gravitational perturbations}. However, to a
good approximation, QNM frequencies of the DF equation for $s=-1$ are
{\it very close} to frequencies obtained freezing metric perturbations
in RN when we consider {\it small} values of the charge, say
$Q\lesssim M/2$. This is also true for EM QNM frequencies with $l=2$,
in which case EM perturbations are coupled with gravitational
perturbations.

\item[iii)] Inspection of gravitational QNMs with $l=2$ confirms the
above conclusions: {\it when we consider the DF equation with $s=-2$
we are not really ``killing'' EM perturbations}. However, to a good
approximation, QNM frequencies of the DF equation for $s=-2$ are {\it
very close} to those obtained freezing EM perturbations in RN when we
consider {\it small} values of the charge, say $Q\lesssim M/2$.

\item[iv)] For RN black holes with $Q\lesssim M/2$ the DF equation
provides the correct RN oscillation frequencies of the full {\it
coupled} EM-gravitational perturbations system within about $1~\%$. We
can reasonably expect our results to have the same level of accuracy
for rotating, KN black holes, at least when $Q\lesssim M/2$ and $a$ is
not too large.

\end{itemize}

\subsection{Asymptotic modes of ``frozen'' RN black holes}

Highly damped black hole QNMs received considerable attention
recently, due to a conjectured relation with quantum gravity
\cite{HD}. Motl and Neitzke \cite{MN} used a monodromy calculation to
compute asymptotic QNM frequencies of RN black holes. The key element
of their calculation is the leading asymptotic behavior of the
potential close to the origin. They showed that, since the leading
term of both potentials in (\ref{axialcoupled}) as $r\to 0$ is
\be
V_i\sim \f{j^2-1}{4r_*^2}\,,\qquad (i=1,2)\,,
\ee
with $j=5/3$, asymptotic QNMs are given by the implicit formula
\be\label{MNf}
e^{\beta \omega}+2+3e^{-\beta_I \omega}=0\,.
\ee 
where $\beta$ and $\beta_I$ are the Hawking temperatures of the outer
and inner RN horizons, respectively \cite{MN}. The ``frozen''
potentials (\ref{dec}) have the same leading-order behavior:
\be
V^*_i\sim \f{4Q^4}{r^6}\sim \f{4}{9r_*^2}\,,\qquad (i=1,2)\,,
\ee
where we used the fact that $r_*\sim r^3/(3Q^2)$ as $r\to 0$.
Therefore, if we kill EM (or gravitational) perturbations {\it keeping
the charge in the background}, the asymptotic QNMs are still given by
(\ref{MNf}). In this sense, the wild oscillations of asymptotic QNM
frequencies as a function of $Q$ are {\it not} induced by the
EM-gravitational coupling. In fact, the presence of the charge term in
the potential (and {\it not} the coupling of EM-gravitational
perturbations) is the reason for the different topology of the Stokes
lines in the monodromy calculation \cite{AH}.

\section{Conclusions}
\label{sec:conc}

In this paper we presented the first continued-fraction calculation of
QNMs of the KN black hole based on the approximate perturbation
equation derived by Dudley and Finley \cite{DF}. For scalar
perturbations their equation is exact; for EM and gravitational
perturbations, it provides a good approximation for values of the
charge $Q\lesssim M/2$. We found no evidence for instabilities,
extending a previous analysis by Mashhoon \cite{Mash}, who presented
arguments in favor of the stability of KN black holes in the eikonal
approximation. To understand the meaning of the DF approximation we
analysed in detail the zero-rotation (RN) limit. We computed QNMs for
``frozen'' (purely EM or gravitational) perturbations of RN black
holes and compared results with the zero-rotation limit of the DF
equation. We found that the DF equation is equivalent to frozen EM
(gravitational) perturbations of RN only when $Q\lesssim M/2$. In this
regime, and for $a=0$, the DF equation provides the correct QNMs of
the full {\it coupled} EM-gravitational perturbations system within
about $1~\%$. We expect our results to have the same level of accuracy
also for rotating black holes, at least when $Q\lesssim M/2$ and $a$
is not too large.

To confirm this expectation we need a more complete analysis of the KN
perturbation problem. Some suggestions to decouple EM and
gravitational perturbations of KN black holes were given in \cite{BF}
and, more recently, in \cite{PV}. Chitre \cite{chitre} found a
separable equation for rotating black holes with small values of
$Q$. His treatment was later extended by Lee \cite{lee} (see also
\cite{CC}). As far as we know, no attempt has been made to compute the
corresponding QNMs. Results obtained from Lee's small-$Q$ expansion
should not deviate much from ours, since the DF equation gives
accurate QNM frequencies in the limit $Q\ll M$.

Our study shows that the DF equation can be used to provide reliable
estimates of the QNM frequencies of slightly charged, rotating BHs.
Mashhoon's approximate treatment, summarized in the Appendix, captures
the essential physics and gives very good estimates for the $l=|m|=2$
fundamental QNM as a function of both $Q$ and $a$ (see
Fig.~\ref{fig:mash}). These results could be useful to investigate
gravitational wave emission in coincidence with black hole models of
gamma-ray bursts. Proposed models to date include bipolar outflow from
a fast-rotating, slightly charged KN black hole surrounded by a
magnetosphere \cite{P}, black hole-torus systems \cite{VP}, pair
creation due to vacuum polarization in gravitational collapse to a
charged black hole \cite{ruffini} and hyper-accretion scenarios
\cite{AG}.

An interesting extension of our work concerns the study of Dirac
perturbations of the KN metric. In this case separability is not an
issue \cite{page}. Pekeris and collaborators considered the {\it
nucleus} as a Kerr-Newman source \cite{pekeris}. Assuming the angular
momentum of the source to be the intrinsic spin angular momentum of
the nucleus they found a remarkable result: the Dirac equation in the
KN background predicts the hyperfine splitting observed in muonium,
positronium and hydrogen to within the uncertainty in the respective
QED corrections, {\it except for an unaccounted factor of 2}.  Their
analysis is worth revisiting, considering also that a very complete
mathematical study of the Dirac equation in the KN spacetime appeared
after Pekeris' work \cite{KM}. We hope to return to this problem in
the future.

\section*{Acknowledgements}

We are grateful to Nils Andersson, Vitor Cardoso and Christian
Cherubini for useful discussions, to Jim Meyer for a careful reading
of the manuscript and to Bahram Mashhoon for bringing to our attention
Ref.~\cite{Mash}. This work was supported in part by the National
Science Foundation under grant PHY 03-53180.

\appendix
\section{Perturbed orbits of null rays and the eikonal limit}\label{mashhoon}

In this Appendix we summarize Mashhoon's analysis of the perturbations
of unstable circular photon orbits in the KN spacetime \cite{Mash}.
His physical model is closely related to (but slightly different from)
Goebel's \cite{goebel}. The key idea is that, since QNMs represent a
general property of the spacetime, we can find them using the time
evolution of any convenient perturbation. In particular, Mashhoon
considers an aggregate of massless particles in the unstable
equatorial circular orbit of a KN black hole. The radius $r_0$ of this
orbit is given by the roots of
\be\label{UPO}
r_0^2-3r_0/2+2Q^2\pm2a(r_0/2-Q^2)^{1/2}=0\,.
\ee
The upper sign refers to corotating and the lower sign to
counterrotating orbits, and only solutions with $r_0\geq r_+$ are
acceptable. The corresponding orbital frequencies are
\be
\omega_{\pm}=\f{1}{a\pm r_0^2(r_0/2-Q^2)^{-1/2}}\,.
\ee
According to Mashhoon's model, the real part of QNM frequencies in the
eikonal limit is proportional to the frequency of the perturbed bundle
of null rays that escape to infinity, $\omega_{\pm}$, and the
imaginary part is proportional to
\be
\gamma=|\omega_{\pm}|\f{4\Delta(r_0)[3(2r_0)^{-1}-4Q^2 r_0^{-2}]^{1/2}}{r_0(2r_0-1)}\,.
\ee
Perturbed null rays corotating (counterrotating) with the black hole
correspond to $l=\pm m$, so Mashhoon's analysis predicts QNM
frequencies
\be\label{mQNM}
\omega^{(M)}=\left(\pm l \omega_{\pm},-(n-1/2)\gamma\right)\,,
\ee
where the $-1/2$ in the imaginary part accounts for the fact that we
count modes starting from $n=1$, following Leaver's convention
\cite{L}.  In the Schwarzschild limit Mashhoon's model predicts
$\omega_{\pm}=\gamma=2/(3\sqrt{3})\simeq 0.384900$, in agreement with
the leading-order large-$l$ expansion from the WKB approximation
(Eq.~(3.1) in \cite{sai}):
\be
\omega=\left(\f{2l}{3\sqrt{3}}, -\f{2(n-1/2)}{3\sqrt{3}}\right)\,.
\ee
In the extremal limit we can eliminate (say) $Q$ in favour of $a$ in
Eq.~(\ref{UPO}) and solve analytically for the extremal frequencies
$\omega^{\rm ext}_{\pm}$. For corotating orbits the result is
$\omega^{\rm ext}_+=(2-3a)^{-1}$ for $0\leq a\leq 1/4$, and
$\omega^{\rm ext}_+=4a(1+4a^2)^{-1}$ for $1/4\leq a\leq 1/2$. For
counterrotating orbits, $\omega^{\rm ext}_-=-(2+3a)^{-1}$ in the whole
range $0\leq a\leq 1/2$.


\begin{thebibliography}{00}

\bibitem{KN}
R. P. Kerr, 
Phys. Rev. Lett. {\bf 11}, 237 (1963);
E. T. Newman, E. Couch, K. Chinnapared, A. Exton, A. Prakash and
R. Torrence, J. Math. Phys. {\bf 6}, 918 (1965).
For a review of black hole uniqueness theorems cf.
M. Heusler, 
Living Rev. Relativ. {\bf 1}, 6 (1998).
\bibitem{MTB} 
S. Chandrasekhar, 
{\it The Mathematical Theory of Black Holes}
(Oxford University, New York, 1983).
\bibitem{QNM} 
K.D. Kokkotas and B.G. Schmidt,
Living Rev. Relativ. {\bf 2}, 2 (1999).
\bibitem{Carter}
B. Carter,
Phys. Rev. {\bf 174}, 1559 (1968).
\bibitem{Rosquist}
K. Rosquist,
gr-qc/0412064.
\bibitem{KNmodels}
O. Gron,
Phys. Rev. D {\bf 31}, 2129 (1985);
T. Ledvinka, M. Zofka and J. Bicak, 
in {\it Proceedings of the Eighth Marcel Grossmann meeting on recent
developments in theoretical and experimental general relativity,
gravitation and relativistic field theories}, edited by T. Piran
(World Scientific, Singapore, 1999), available online as
gr-qc/9801053;
H. I. Arcos and J. G. Pereira, 
Gen. Rel. Grav. {\bf 36}, 2441 (2004).
\bibitem{pekeris}
C. L. Pekeris, 
Phys. Rev. A {\bf 35}, 14 (1987);
C. L. Pekeris and K. Frankowski, 
Phys. Rev. A {\bf 39}, 518 (1989).
\bibitem{BZ}
R. D. Blandford and R. Znajek,
MNRAS {\bf 179}, 433 (1977).
\bibitem{P}
B. Punsly,
Astrophys. J. {\bf 498}, 640 (1998);
B. Punsly,
Astrophys. J. {\bf 498}, 660 (1998).
\bibitem{VP} 
M. H. P. M. van Putten,
Phys. Rev. Lett. {\bf 87}, 091101 (2001);
M. H. P. M. van Putten,
Phys. Rep. {\bf 345}, 1 (2001).
\bibitem{ruffini} 
R. Ruffini, C. L. Bianco, P. Chardonnet, F. Fraschetti, L. Vitagliano
and S.-S. Xue, in AIP Conf. Proc. {\bf 668}, 16 (2003); astro-ph/0302557.
\bibitem{AG}
R. A. Araya-G\'ochez,
MNRAS {\bf 355}, 336 (2004).
\bibitem{shapiro}
C. F. Gammie, S. L. Shapiro and J. C. McKinney,
Astrophys. J. {\bf 602}, 312 (2004).
\bibitem{DT}
N. Dadhich and Z. Y. Turakulov,
Class. Quantum Grav. {\bf 19}, 2765 (2002).
\bibitem{page}
D. N. Page,
Phys. Rev. D {\bf 14}, 1509 (1976).
\bibitem{DF} 
A. L. Dudley and J. D. Finley III,
J. Math. Phys. {\bf 20}, 311 (1979).
\bibitem{DFPRL} 
A. L. Dudley and J. D. Finley III,
Phys. Rev. Lett. {\bf 38}, 1505 (1977).
Phys. Rev. Lett. {\bf 39}, 367E (1977).
\bibitem{PD} J. F. Pleba\'nski and M. Demia\'nski.
Ann. Phys. (N.Y.) {\bf 98}, 98 (1976).
\bibitem{Weir} G. J. Weir, {\it Type D Spaces and
Quasidiagonalizability}, Ph. D Thesis, University of Canterbury,
Christchurch, New Zealand (1976); W. Kinnersley, J. Math. Phys. {\bf
10}, 1195 (1969).
\bibitem{Mash} B. Mashhoon,
Phys. Rev. D {\bf 31}, 290 (1985).
\bibitem{goebel} C. G. Goebel,
Astrophys. J. Lett. {\bf 172}, L95 (1972).
\bibitem{KNC} K.D. Kokkotas,
Nuovo Cimento {\bf 108B}, 991 (1993).
\bibitem{L} E. W. Leaver, 
Proc. Roy. Soc. Lon. {\bf A402}, 285 (1985). 
\bibitem{DO} S. Detweiler and R. Ove,
Phys. Rev. Lett. {\bf 51}, 67 (1983).
\bibitem{teu} S. A. Teukolsky, 
Astrophys. J. {\bf 185}, 635 (1973).
\bibitem{N} H.-P. Nollert, Phys. Rev. D {\bf 47}, 5253 (1993). 
\bibitem{KS}
K. D. Kokkotas, B. F. Schutz,
Phys. Rev. D {\bf 37}, 3378 (1988).
\bibitem{O} 
H. Onozawa,
Phys. Rev. D {\bf 55}, 3593 (1997).
\bibitem{AO} N. Andersson and H. Onozawa, 
Phys. Rev. D {\bf 54}, 7470 (1996).
\bibitem{BK}
E. Berti and K. D. Kokkotas, 
Phys. Rev. D {\bf 68}, 044027 (2003).
\bibitem{BCKO}
E. Berti, V. Cardoso, K. D. Kokkotas and H. Onozawa, 
Phys. Rev. D {\bf 68}, 124018 (2003).
\bibitem{BCY}
E. Berti, V. Cardoso and S. Yoshida,
Phys. Rev. D {\bf 69}, 124018 (2004).
\bibitem{GA} 
K. Glampedakis and N. Andersson,
Class. Quantum Grav. {\bf 20}, 3441 (2003).
\bibitem{milano}
E. Berti,
in {\it Proceedings of the International Workshop on Dynamics and
Thermodynamics of Black Holes and Naked Singularities} (Milan, Italy,
13-15 May 2004); gr-qc/0411025.
\bibitem{De} 
S. Detweiler,
Astrophys. J. {\bf 239}, 292 (1980).
\bibitem{GADet}
K. Glampedakis and N. Andersson,
Phys. Rev. D {\bf 64}, 104021 (2001).
\bibitem{CDet}
V. Cardoso,
Phys. Rev. D {\bf 70}, 127502 (2004).
\bibitem{sai}
S. Iyer,
Phys. Rev. D {\bf 35}, 3632 (1987).
\bibitem{KerrQNM} 
E. Seidel and S. Iyer,
Phys. Rev. D {\bf 41}, 374 (1990); 
K. D. Kokkotas, 
Class. Quantum Grav. {\bf 8}, 2217 (1991).
\bibitem{FHH}
C. L. Fryer, D. E. Holz and S. A. Hughes,
Astrophys. J. {\bf 565}, 430 (2002).
\bibitem{L2} 
E. W. Leaver, 
Phys. Rev. D {\bf 41}, 2986 (1990).
\bibitem{Burko} 
Burko pointed out that the perturbative formalism leading to
Eqs. (\ref{axialcoupled}) fails for $l=1$, and corrected
Chandrasekhar's treatment in this special case: see L. Burko,
Phys. Rev. D {\bf 59}, 084003 (1999) and L. Burko, Phys. Rev. D {\bf
52}, 4518 (1995). His final result for $l=1$ is identical to our
decoupled Eqs. (\ref{dec}).
\bibitem{BKRN}
E. Berti and K. D. Kokkotas,
Phys. Rev. D {\bf 68}, 044027 (2003).
\bibitem{IW}
S. Iyer and C. M. Will,
Phys. Rev. D {\bf 35}, 3621 (1987).
\bibitem{Kon}
R. A. Konoplya,
Phys. Rev. D {\bf 68}, 024018 (2003).
\bibitem{HD} 
S. Hod,
Phys. Rev. Lett. {\bf 81}, 4293 (1998);
O. Dreyer,
Phys. Rev. Lett. {\bf 90}, 081301 (2003).
\bibitem{MN} 
L. Motl and A. Neitzke,
Adv. Theor. Math. Phys. {\bf 7}, 307 (2003).
\bibitem{AH}
N. Andersson and N. Howls,
Class. Quantum Grav. {\bf 21}, 1623 (2004).
\bibitem{BF}
V. Bellezza and V. Ferrari,
J. Math. Phys. {\bf 25}, 1985 (1984).
\bibitem{PV}
Z. Perj\'es and M. Vas\'uth,
Astrophys. J. {\bf 582}, 342 (2003).
\bibitem{chitre}
D. M. Chitre,
Phys. Rev. D {\bf 11}, 760 (1975);
D. M. Chitre,
Phys. Rev. D {\bf 13}, 2713 (1976).
\bibitem{lee}
C. H. Lee,
Prog. Theor. Phys. {\bf 66}, 180 (1981).
\bibitem{CC}
C. Cherubini and R. Ruffini, 
Nuovo Cimento {\bf 115B}, 7 (2000).
\bibitem{KM}
E. G. Kalnins, W. Miller, Jr., 
J. Math. Phys. {\bf 33}, 286 (1992).
\end{thebibliography}
\end{document}